# The Nonequilibrium Nature of Culinary Evolution

Osame Kinouchi[1], Rosa W. Diez-Garcia[2], Adriano J. Holanda[1], Pedro Zambianchi[3] and Antonio C. Roque[1]

[1]*Departamento de Física e Matemática, FFCLRP, Universidade de São Paulo, 14040-901 Ribeirão Preto, SP, Brazil*

[2]*Departamento de Clínica Médica, Curso de Nutrição e Metabolismo, Faculdade de Medicina de Ribeirão Preto, Universidade de São Paulo, 14040-901 Ribeirão Preto, SP, Brazil*

[3]*Faculdade Bandeirantes, 14010-060 Ribeirão Preto, SP, Brazil*

Food is an essential part of civilization, with a scope that ranges from the biological to the economic and cultural levels. Here we study the statistics of ingredients and recipes taken from Brazilian, British, French, and Medieval cookbooks. We find universal distributions with scale invariant behavior. We propose a copy-mutate process to model culinary evolution that fits very well our empirical data. We find a cultural "founder effect" produced by the nonequilibrium dynamics of the model. Both the invariant and idiosyncratic aspects of culture are accounted by our model, which may have applications in other kinds of evolutionary processes.

Almost every human group has its own cuisine [1–2], usually one of the dearest aspects of its cultural identity. The diversity of foodstuffs that constitute a cuisine as well as their social and symbolic aspects are of permanent concern to anthropologists, psychologists and sociologists [3]. From another perspective, the chemical-physics of culinary processes has attracted the interest of physicists and chemists [4–5].

Here we propose a statistical [6] and complex networks [7] approach to model culinary diversity and evolution. Culinary recipes are prime examples of cultural algorithms with strong capacity for stabilization, innovation and transmission. In literate cultures their standardized reproduction and widespread use are favored by cookbooks, which represent their culinary traditions. Cookbooks provide statistical information about cuisines, indicating the relative importance of foodstuffs, preparation ways and combinations of them in a cuisine. These elements have made traditional cookbooks a useful source of information for food researchers [1, 8–9].

In this work we examine only a particular aspect of cookbooks, namely the relationship between its recipes and ingredients, and neglect for the moment the very important role played by culinary preparations. We study the statistics of ingredients usage in different countries and cultures in search for common statistical patterns or differences between them.

We also propose a copy-mutate algorithm to model cuisine growth from a few initial recipes, which fits very well our empirical data. The model suggests an evolutionary dynamics where idiosyncratic ingredients are preserved in a manner akin to the founder effect in biology [10].

The culinary corpus considered by us consists of four different cookbooks: the Brazilian *Dona Benta* [11], the French *Larousse Gastronomique* [12], the British *New Penguin Cookery Book* [13], and the medieval *Pleyn Delit* [14]. We have constructed a database for each one of them containing its entire population of recipes and ingredients. The only exception is the database for the French cookbook, which contains a random sample (40%) of the whole population of recipes. To examine temporal effects, we have considered three very different editions of *Dona Benta* [11] (1946, 1969 and 2004) which largely differs in recipes and ingredients repertoire. The numbers of recipes and ingredients in each database are given in Table 1.

For each cookbook database, we counted the number of times each ingredient appeared in the recipes and ordered them according to descending frequency. This allows the representation of the statistical hierarchy of ingredients in a cookbook by a rank-frequency plot, as introduced by Zipf [15]. Fig. 1a gives the rank-frequency plots for the Brazilian (1969 ed.), British, French and medieval cookbooks, showing a remarkable similarity in their statistical patterns.

Time-invariance is supported by the data shown in Fig. 1b, which give the rank-frequency plots for the editions of the Brazilian cookbook. The structure of ingredient rankage in the Brazilian cookbook remained stable amidst the change from a regional to a more globalized food consumer profile that took place in the last fifty years.

All these curves exhibit a power-law behavior which can be well fitted by a Zipf-Mandelbrot law [16] with an exponential cut-off [17] to capture finite size effects,

$$f(r) = \frac{C}{a + r^\beta} \exp(-r/r_c). \qquad (1)$$

The best fit exponent $\beta \cong 1.4$ worked reasonably well for all curves suggesting that, like Zipf's law in linguistics [15], there exists a statistical universal behavior which is robust across different cookbooks and independent of the culture they refer to, their authors, motivations and even time.

A common measure used to characterize complex networks is the degree distribution, which gives the fraction of nodes with $k$ links [7]. A bipartite network has two kinds of degree distribution, one for the ingredient nodes and the other for the recipe nodes. The degree distribution for the ingredients is the probability distribution $P^I(k)$ that a randomly chosen ingredient appears in $k$ recipes. Plots of the degree distribution for the ingredients of the cookbooks show that they are right-skewed with a two decade power law region well fitted by $P^I(k) \propto k^{-\alpha}$ with $\alpha \cong 1.72$. This value has been found by a fit proportional to $k^{-(\alpha-1)}$ of the complementary cumulative distribution (Fig. 2),

$$P_C(k) = 1 - \sum_k P^I(k) , \qquad (2)$$

which gives the probability that an ingredient is used in more than $k$ recipes. It is compatible with the relation $\alpha = 1 + 1/\beta$ valid for pure power laws present in rank plots and degree distributions. This anomalous exponent cannot be obtained from a general Yule process [6] since in this latter case $\alpha \geq 2$.

The degree distribution for the recipes is the probability distribution $P^R(j)$ that a randomly chosen recipe has $j$ ingredients (what we will call the size $j$ of the recipe). A plot of the degree distribution for the recipes of the French cookbook is shown in Fig. 2 (inset). The average recipe sizes for all cookbooks are given in Table 1.

On the basis that culinary recipes are examples of cultural replicators, or "memes" [18], we propose a copy-mutate algorithm to model culinary evolution as a branching process (Fig. 3 inset). Our attempts to fit all the frequency distributions have shown that our model needs at least five parameters to work: the number $T$ of generations (iterations or "time") until the process is halted, the number $K$ of ingredients per recipe (where $K$ is compared to the average number $<K>$ of the cookbook), the number $L$ ($\neq K$) of ingredients ("loci") in each recipe to be mutated, the number $R_0$ of initial recipes in the cuisine and $M$, the ratio between the sizes of the pool of ingredients and the pool of recipes. Hence, starting with $t_0 = R_0$, at each time step there are $R(t) = t$ recipes and $MR(t)$ ingredients available to be used.

We tried models without the concept of fitness, with no success. So, the present model ascribes to each ingredient $i$ a random fitness $f_i$ with values uniformly distributed in the interval *[0,1]*. We interpret this fitness as related to intrinsic ingredient properties as nutritional value, aspect, flavor, cost and availability. At each iteration, we randomly choose a recipe (a "mother") and copy it. Within this copy we randomly choose an ingredient (with fitness $f_i$) to be compared with an ingredient also randomly chosen from the ingredients pool: if $f_j > f_i$, where $f_j$ is the fitness of the ingredient from the pool, we replace ingredient $i$ by ingredient $j$. This process is repeated $L$ times, thus generating a "daughter" recipe that is added to the recipes pool (it is possible that the daughter remains identical with the mother, which we interpret as a new recipe that differs from the previous one in the cooking procedures but not in their ingredients). Finally, at each time step, new ingredients are introduced in the ingredients pool to maintain the ratio $M$ fixed. A somewhat similar algorithm has been proposed by Ramezanpour [19], but without the introduction of (the very important) fitness-based selection process.

A more general fitness function could be used,

$$f_i^I = f_i + \sum_j f_{ij} + \sum_{j,k} f_{ijk} + \ldots,$$

to account for higher order interactions between ingredients, which would mean that the fitness of an ingredient is contextual and depends on the other ingredients present in the recipe.

We searched in the parameter space $(T, K, L, M, R_0)$ for a good fit for the rank-frequency plot of the Larousse cookbook (Fig. 3), and obtained the values $T = 1200$, $K = 11$, $L = 4$, $M = 3$ and $R_0 = 20$. They show a good agreement with the actual values of the Larousse database, since the number of recipes is $R = T = 1200 = R_{\text{Larousse}}$ and $K = 11 \cong <K>_{\text{Larousse}} = 10.8$. So, the effective free parameters are only the initial number of recipes $R_0$ (or independent "filogenetic trees"), the mutation rate $L$ and the growth rate $M$ of the ingredients pool, which are not determinable from empirical data. Parameter fitting also gives a good agreement with the curves and values for all the other cookbooks considered.

We also define the fitness of the $k$th recipe as $F^{(k)} = \frac{1}{K}\sum_{i=1}^{K} f_i$ and a total time dependent cuisine fitness $F_{\text{Total}}(R(t)) = \frac{1}{R(t)}\sum_{k=1}^{R(t)} F^{(k)}$. In Fig. 4, we examine the temporal evolution of $1 - F_{\text{Total}}(t)$, which shows a very slow convergence to equilibrium in the form of a power law $1 - F_{\text{Total}}(t) \propto t^{-\gamma}$, with $\gamma = 0.1$. Hence, this kind of historical dynamics has a glassy character, where memory of the initial conditions is preserved, suggesting that the idiosyncratic nature of each cuisine will never disappear due to invasion by alien ingredients and recipes.

It is interesting that the same model presents a "stationary" state (the rank-frequency power law) coexisting with a power law convergence toward a global fitness maximum. We conjecture that this scale invariance arises because the model implements a critical branching process [20]: the branching ratio is $\sigma = R(t+1)/R(t) \cong 1 = \sigma_c$.

The evolutionary model of copy-mutation of recipes along with a selection mechanism generates a scale-free cuisine. This evolution is an out-of-equilibrium process: the number of recipes is never sufficient to fully explore the combinatorial space of ingredients. This means that the invasion of new high fitness ingredients and the elimination of initial low fitness ingredients never end. The latter is related to the "founder effect" [16] known in evolutionary theory: in spite of their low fitness some of the $KR_0$ ingredients present in the initial $R_0$ recipes have a strong difficulty of being probabilistically replaced. They are like frozen "cultural" accidents difficult to be overcome in the out-of-equilibrium regime (see Fig. 5).

We notice that sometimes in biology gene fitness is taken as proportional to the relative frequency in the genome. However, this would occur only in an equilibrium state. Our model produces a stationary out-of-equilibrium state (the power law) allowing the appearance of the above "founder effect"-like phenomenon.

In our simulations we have found that Darwinian selection by ingredient fitness is essential to obtain good modeling for rank values in the interval $r = 1,\ldots, KR_0$. We tested an algorithm without fitness selection and observed that the conservation of the first $KR_0$ ingredients (the founder effect) is so strong that they remain over represented in the population.

Our model is a general copy-mutate mechanism with fitness applicable to evolutionary processes, of biological or cultural nature, like culinary growth. We found that both universal (scale invariance) and particular (founder effect) properties of evolutionary phenomena can be accounted for by a single nonequilibrium critical dynamics.

Acknowledgments

We thank N. Caticha, M. Copelli, A. S. Martinez and P. Riegler for discussions and encouragement, and M. A. Bellezi, R. F. Caron, J. G. Costa, D. M. Gonçalves, D. G. Lidovero, G. Marini, S. T. Pinto and S. M. F. Ribeiro for preliminary statistical results O.K. and A. R. are supported by research grants from CNPq.

References

1. J. L. Flandrin, M. Montanari and A. Sonnenfeld (eds.), *Food: A Culinary History* (Penguin, New York, 2000).

2. R. MacLennan and A. Zhang, Asia Pacific J. Clin. Nutr. **13**, 131 (2004).

3. M. Harris and E. B. Ross (eds.), *Food and Evolution: toward a theory of human food habits* (Temple University Press, Philadelphia, PA, 1987).

4. G. Weiss, Science **293**, 1753 (2001).

5. H. This, Nature Mater. **4**, 5 (2005).

6. M. E. J. Newman, Contemp. Phys. **46**, 323 (2005).

7. M. E. J. Newman, A. L. Barabási, and D. J. Watts, *The Structure and Dynamics of Networks* (Princeton University Press, Princeton, 2006).

8. A. Appadurai, Comp. Stud. Soc. Hist. **30**, 3 (1988).

9. Y. Segers, Appetite **45**, 4 (2005).

10. E. Mayr, *Animal Species and Evolution* (Harvard University Press, Cambridge, MA, 1963).

11. *Dona Benta: Comer Bem*. (26th ed., 1946; 51st ed. rev., 1969; 76th ed. rev., 2004) (Companhia Editora Nacional, São Paulo).

12. *Larousse Gastronomique* (Larousse, Paris, 2004).

13. J. Norman, *The New Penguin Cookery Book*. (Penguin, London, 2004).

14. C. B. Hieatt, B. Hosington and S. Butler, *Pleyn Delit: medieval cookery for modern cooks*. 2nd ed. (University of Toronto Press, Toronto, 1996).

15. G. K. Zipf, *Human Behavior and the Principle of Least Effort: an Introduction to Human Ecology*. (Addison-Wesley, Cambridge, MA, 1949).

16. B. B. Mandelbrot, *The Fractal Geometry of Nature*. (Freeman, New York, 1977).

17. L. A. N. Amaral, A. Scala, M. Barthélémy and H. E. Stanley, Proc. Natl. Acad. Sci. USA **97**, 11149 (2000).

18. R. Dawkins, *The Selfish Gene*. 30th Anniversary Edition. (Oxford University Press, Oxford, 2006).

19. A. Ramezanpour, Europhys. Lett. **68**, 316 (2004).

20. O. Kinouchi and M. Copelli, Nature Phys. **2**, 348 (2006).

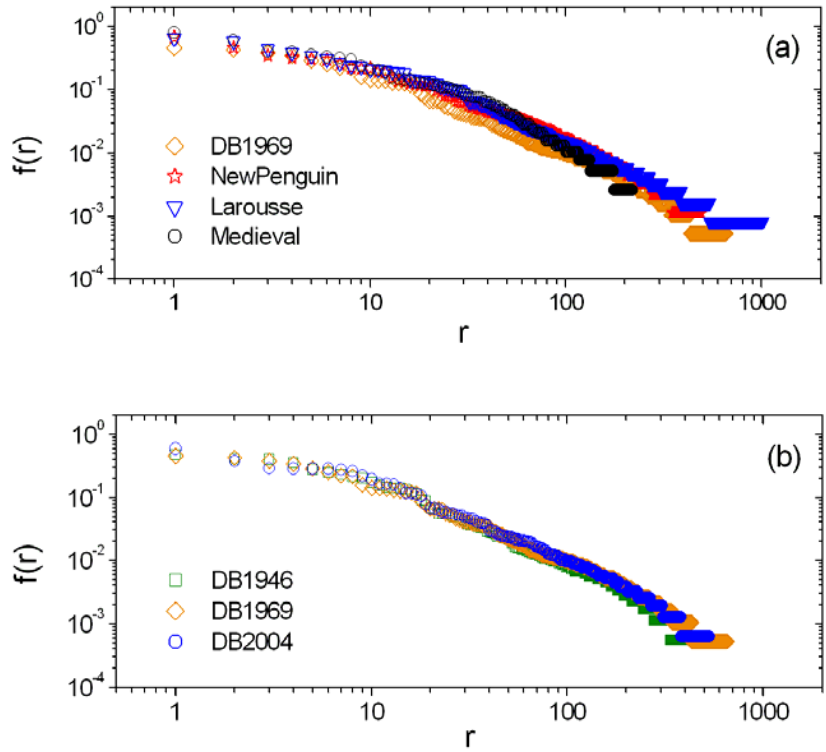

FIG. 1 (color online). (a) Cultural invariance. Rank-frequency plot for different cookbooks: *Pleyn Delit* (circles), *Dona Benta* 1969 (squares), *New Penguin* (stars) and *Larousse Gastronomique* (triangles). (b) Temporal invariance. Rank-frequency plot for *Dona Benta* 1946 (squares), 1969 (lozenge) and 2004 (circles).

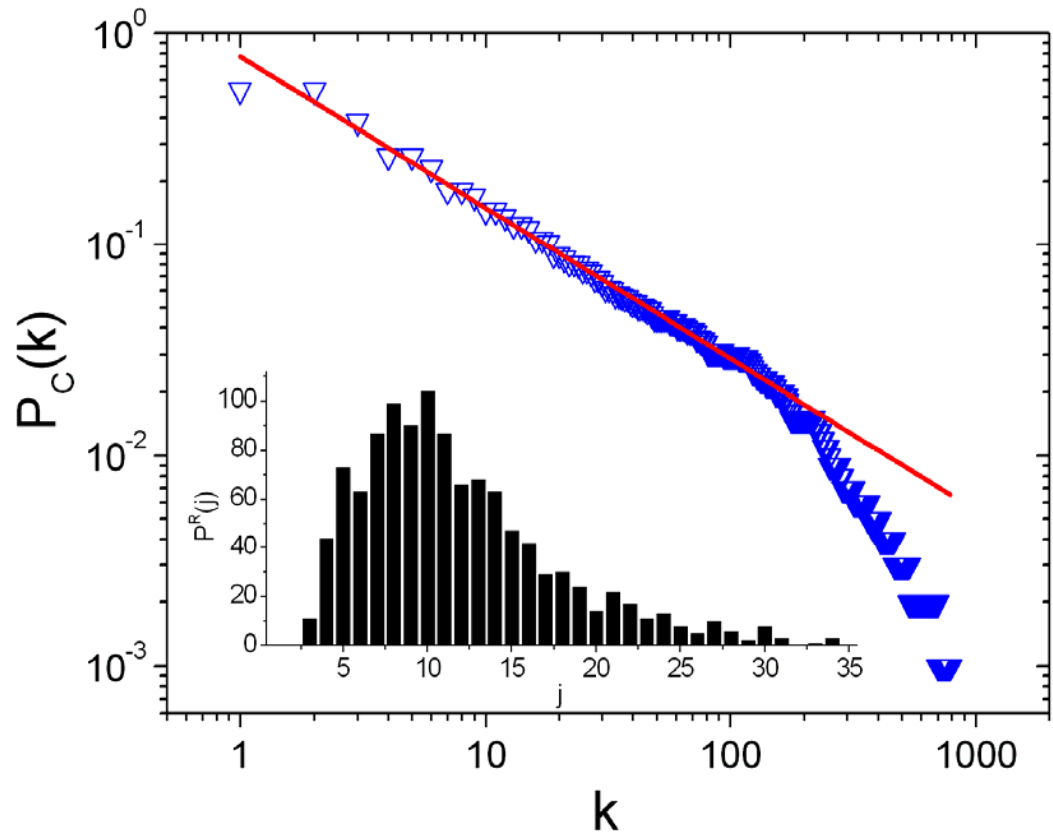

FIG. 2 (color online). Complementary cumulative degree distribution $P_C(k)$ for ingredients with power law fitting $k^{-0.72}$ (solid line). Inset: Degree distribution $P^R(j)$ for recipes.

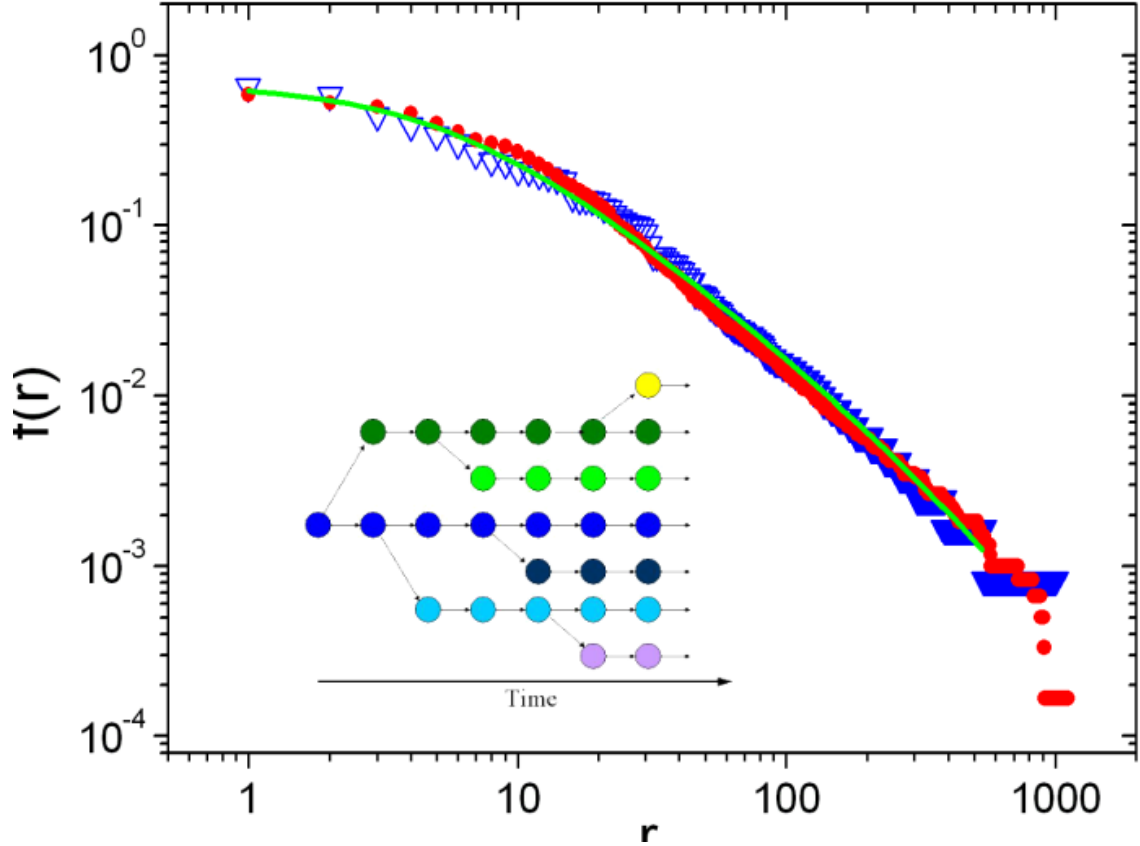

FIG. 3 (color online). Frequency-rank plot for the Laurousse cookbook (triangles) fitted by a Zipf-Mandelbrot curve (solid) and by the model (circles, average of five runs). Inset: Schematic view of the copy-mutate growth process. At each time step, a randomly chosen "mother" recipe generates a mutated daughter recipe. Notice that there is no recipe extinction.

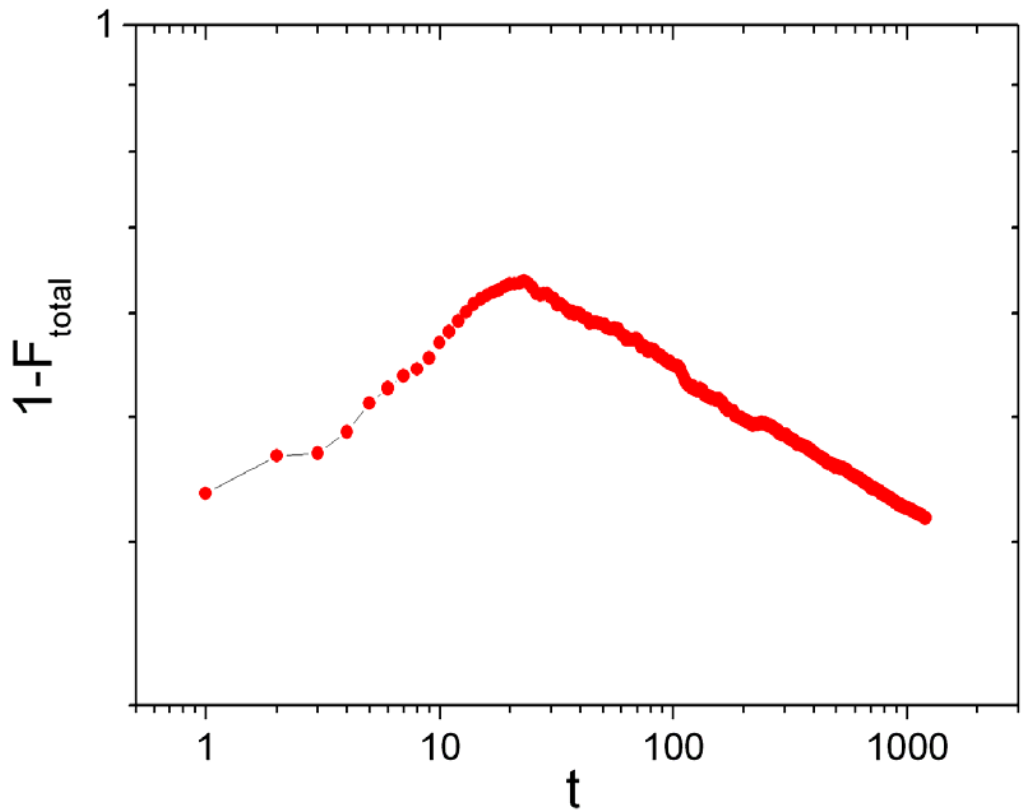

FIG. 4. Power law decay with small exponent γ = − 0.1 observed in the convergence of total culinary fitness $F_{total}$ to 1. The rising phase is due to the initial copy of low fitness recipes, which lowers the overall culinary fitness $F_{total}$.

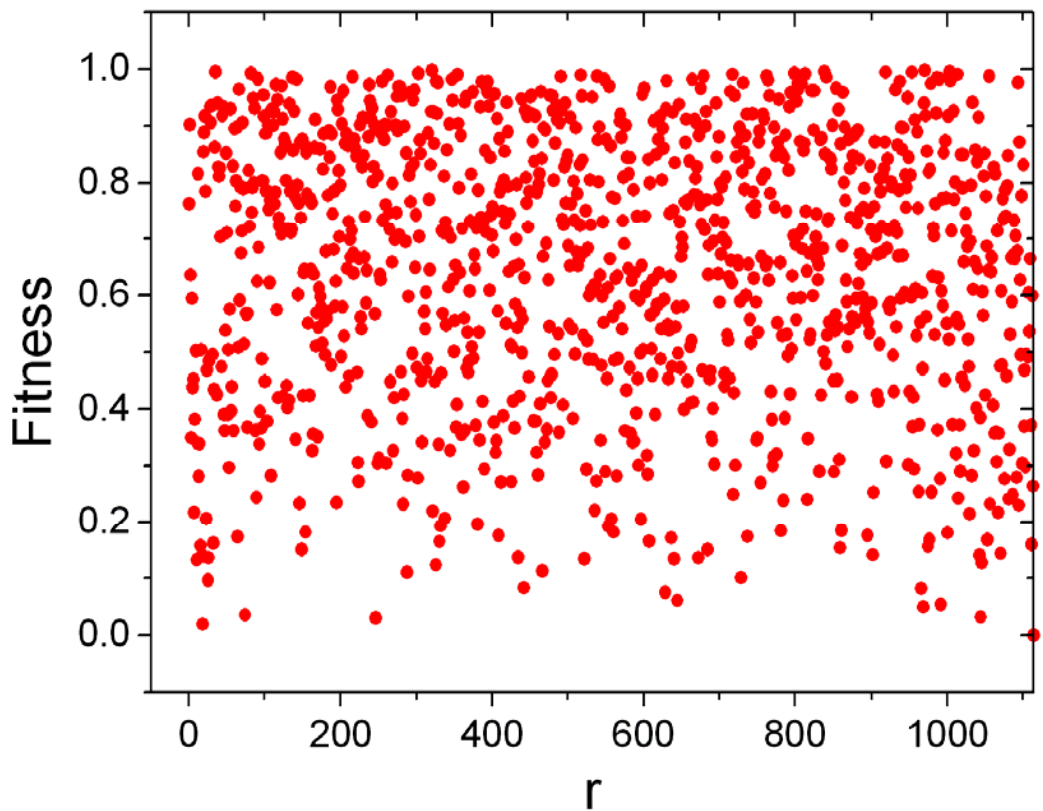

FIG. 5. Scatter plot of ingredient fitness versus ingredient rank *r*. Observe the founder effect for some well ranked ingredients with small fitness (bottom left) and the overall average high fitness due to selective pressure. Notice also that there are several high fitness ingredients with poor ranking (upper right) due to the strong non-equilibrium character of the evolutionary process.

**Table 1. Numerical characteristics for the considered cookbooks.**

| Cookbook | No. of recipes | No. of ingredients | Average recipe size |
| --- | --- | --- | --- |
| Dona Benta (1946) | 1786 | 491 | 6.7 |
| Dona Benta (1969) | 1894 | 657 | 7.1 |
| Dona Benta (2004) | 1564 | 533 | 7.4 |
| Larousse Gastronomique (2004)[*] | 1200 | 1005 | 10.8 |
| New Penguin Cookery Book (2001) | 878 | 489 | 9.5 |
| Pleyn Delit (Medieval) | 380 | 219 | 9.7 |

* Recipes have been sampled. The total number of recipes is above 3000.